\documentclass[prx,amsmath,amssymb,notitlepage,onecolumn,superscriptaddress,longbibliography]{revtex4-2}

\pdfoutput=1

\usepackage{amsmath}
\usepackage{tabularx,graphicx}
\usepackage{epstopdf}
\usepackage{graphicx}
\usepackage{latexsym}
\usepackage{amssymb}
\usepackage{amsmath}
\usepackage{color, colortbl}
\usepackage{psfrag}
\usepackage{bbm}
\usepackage{bm}
\usepackage{titlesec}
\usepackage{dsfont}
\usepackage{feynmp}
\usepackage{slashed}
\usepackage{multirow}

\usepackage[tight]{subfigure}

\usepackage[papersize={8.5in,11in}]{geometry}

\usepackage{color}
\definecolor{darkblue}{rgb}{0.,0.,0.4}
\definecolor{darkred}{rgb}{0.5,0.,0.}
\definecolor{BlueViolet}{RGB}{138,43,226}
\definecolor{SkyBlue}{RGB}{30,144,255}
\definecolor{DarkGreen}{RGB}{0,100,0}
\usepackage[pdftex,colorlinks=true,linkcolor=darkblue,citecolor=blue,urlcolor=blue]{hyperref}

\def\be{\begin{eqnarray}}
\def\ee{\end{eqnarray}}
\def \be{\begin{equation}}
\def \ee{\end{equation}}
\def \bea{\begin{eqnarray}}
\def \eea{\end{eqnarray}}
\def \nn{\nonumber \\}

\begin{document}

\title{Thermoelectric and thermal properties of the weakly disordered non-Fermi liquid phase of Luttinger semimetals}

\author{Hermann Freire}
\affiliation{Instituto de F{\'i}sica, Universidade Federal de Goi{\'a}s, 74.001-970,
Goi{\^a}nia-GO, Brazil}

\author{Ipsita Mandal}
\affiliation{Institute of Nuclear Physics, Polish Academy of Sciences, 31-342 Krak\'{o}w, Poland}
\affiliation
{Department of Physics, Stockholm University, AlbaNova University Center, 106 91 Stockholm, Sweden}

\begin{abstract}
We compute the thermoelectric and thermal transport coefficients in the weakly disordered non-Fermi liquid phase of the Luttinger semimetals at zero doping, where the decay rate associated with the (strong) Coulomb interactions is much larger than the electron-impurity scattering rate. To this end, we implement the Mori-Zwanzig memory matrix method, that does not rely on the existence of long-lived quasiparticles in the system. We find that the thermal conductivity at zero electric field scales as $\bar{\kappa}\sim T^{-n}$ (with $0\lesssim n\lesssim 1)$ at low temperatures, whereas the thermoelectric coefficient has the temperature dependence given by $\alpha\sim T^{p}$ (with $1/2\lesssim p\lesssim 3/2)$.
These unconventional properties turn out to be key signatures of this long sought-after non-Fermi liquid state in the Luttinger semimetals, which is expected to emerge in strongly correlated spin-orbit coupled materials like the pyrochlore iridates. Finally, our results indicate that these materials might be good candidates for achieving high figure-of-merit for thermoelectric applications.
\end{abstract}

\maketitle

\section{Introduction}

Luttinger semimetals arise in strongly correlated spin-orbit coupled materials, that exhibit a Fermi node at the $\Gamma$-point in the Brillouin zone. This node exhibits a quadratic band-touching (QBT) of doubly-degenerate valence and conduction bands in the three-dimensional Brillouin zone. The unconventional properties of these compounds that emerge due to the nontrivial topology of their electronic states, have attracted considerable attention in recent years in the field of topological quantum matter \cite{Kane_RMP,Zhang_RMP,Vishwanath_Weyl,moon-xu,Kondo_2015,armitage_expt,Ohtsuki,rahul-sid,ips-rahul,*ips-rahul2,lukas-herbut,igor16,ips-qbt-sc,ips_qbt_plasmons,ips_qbt_tunnel,Roy_PRB,Herbut-PRB,Pramanik,freire_2020,Witczak-Krempa,ipsfloquet}. Examples of these materials include some pyrochlore iridates \cite{Witczak-Krempa,Kondo_2015}, half-Heusler compounds \cite{Paglione,Taillefer_2013}, and grey-Sn \cite{Groves}, among others.

From a theoretical point of view, the interactions mediated in a Luttinger semimetal by the Coulomb forces provide a key example of a non-Fermi liquid (NFL) state. A controlled theory of this phase can be formulated using a dimensional regularization scheme in $d=4-\varepsilon$ spatial dimensions \cite{Abrikosov_Beneslavskii,Abrikosov}, and it is often dubbed as the Luttinger-Abrikosov-Beneslavskii (LAB) phase. Later on, this classic work was revisited and analyzed by renormalization group (RG) techniques by Moon \emph{et al.} \cite{moon-xu}. They predicted that this NFL phase can in fact be viewed as a ``parent state'' for other novel interesting topological states of matter (which include Weyl semimetal \cite{Vishwanath_Weyl} phases and also a topological insulator \cite{Kane_RMP,Zhang_RMP}), as a result of the application of an external magnetic field and/or strain \cite{Ohtsuki}. We would like to emphasize that the controlled theory of the LAB phase represents an NFL fixed point at a Fermi node, rather than over a Fermi surface \cite{nayak,nayak1,lawler1,mross,Jiang,ips2,ips3,Shouvik1,Freire_RG_PDW,Lee-Dalid,shouvik2,Freire_Pepin_1,ips-uv-ir1,ips-uv-ir2,Freire_Pepin_2,ips-subir,ips-sc,ips-c2,Lee_2018,ips-fflo,ips-nfl-u1}.

A recent study of the transport properties of these systems by Link {\it et al.} \cite{Herbut-PRB} has found that the LAB phase may realize the so-called ``minimal-viscosity'' scenario, in which the ratio of the shear viscosity $\eta$ to the entropy density $s$ is close to the Kovtun-Son-Starinets ratio \cite{DTSon-PRL_2005} $\eta/s \sim 1/(4\pi)$. This condition implies that the complicated quantum dynamics of these systems represent an example of a strongly interacting ``nearly-perfect fluid'', also found in systems like graphene at charge neutrality \cite{Fritz-PRB}, quark-gluon plasma \cite{DTSon-PRL} generated in relativistic heavy-ion colliders, and ultracold quantum gases tuned to the unitarity limit \cite{Cao}. In order to characterize the LAB phase more deeply, it is important to compute other transport coefficients associated with these systems. 

In this paper, we will focus on calculating two central transport coefficients: the thermal conductivity and the thermoelectric response (see Refs~\cite{Muller_2008,Lundgren_2014,Xie_2016,Ghahari_2016,Mandal_Weyl} for such computations in the context of other closely-related systems). The thermal conductivity is naturally related to the ability of a material to conduct heat under an applied temperature gradient, while the thermoelectric response describes the resulting voltage generation due to a temperature gradient. We point out that materials with high thermoelectric efficiency are extremely desirable nowadays for many applications, ranging from power generation to waste heat recovery \cite{Markov_2018}. The conventional wisdom in this area is that the best thermoelectric materials are usually given by heavily-doped semiconductors \cite{Markov_2018}. We will show here that Luttinger semimetals might also be a viable candidate that can have a high thermoelectric response, provided some key conditions are satisfied.

Physically speaking, since the Luttinger semimetals only display a Fermi node in the Brillouin zone center, the electron-electron interactions are not effectively screened in these systems. Consequently, the minimal model to describe these materials must necessarily include strong (long-ranged) Coulomb interactions \cite{moon-xu,rahul-sid,ips-rahul,ips-rahul2}. In addition, due to the absence of long-lived electronic quasiparticles in the LAB phase, the decay rate associated with the Coulomb interactions is expected to be much higher than both electron-impurity and electron-phonon scattering rates. To this end, we will use here an alternative approach to describe the dynamics of these systems, namely the memory matrix formalism, which does not rely on the existence of long-lived quasiparticles \cite{Forster-HFBSCF,Rosch-PRL,Hartnoll-PRB_2013,Freire-AP_2014,Patel-PRB,Hartnoll_PRB_2014,Zaanen-CUP,Lucas-PRB,Freire-AP_2017,Freire-EPL,Sachdev-MIT,Freire-EPL_2018,Berg-PRB,Freire-AP_2020,wang2020low}. The memory matrix method has proved to be a very useful tool to study the hydrodynamic regime in NFL systems. Hence, this framework is tailor-made to address the non-quasiparticle transport regime that emerges in the LAB phase.

The paper is organized as follows. In Sec.~\ref{model}, we define the Hamiltonian of the Luttinger semimetal coupled with long-range Coulomb interactions. Furthermore, we briefly review the main results of the RG analysis, and describe the low-energy NFL fixed point describing the LAB phase. In Sec.~\ref{memory_matrix}, we introduce the memory matrix approach. In Sec.~\ref{results}, we calculate the thermal conductivity and the thermoelectric response of the LAB phase as functions of temperature, in the presence of weak short-ranged disorder. Finally, in Sec.~\ref{conclusions}, we end with the summary and some outlook of our results. Appendix~\ref{leading_corrections} provides the details for the evaluation of the two-loop corrections to the thermal current-momentum susceptibility.
In Appendix~\ref{estimate}, we provide an analysis of the temperature dependence of all transport coefficients, and also of the figure-of-merit, within a simple approximation.

\section{Model}
\label{model}

Our starting point is the Luttinger Hamiltonian \cite{Luttinger}
\begin{equation}\label{bare2}
 \mathcal{H}_0 = \frac{\mathbf k^2}{2 m'}-\frac{\frac{5\,\mathbf k^2}{4}
 -\left (\mathbf{k}\cdot \boldsymbol{\mathcal{J}} \right )^2}{2m}\,,
\end{equation}
where the three components of $ \boldsymbol{\mathcal{J}}$ are the angular momentum operators for $j=3/2$ states. This effective model is known to emerge from the electronic structure of some spin-orbit coupled systems, which display quadratic band-crossings at the Brillouin zone center in three spatial dimensions. It can also
be conveniently recast \cite{moon-xu,rahul-sid,ips-rahul,*ips-rahul,freire_2020} into the form
\begin{equation}
\label{bare}
 \mathcal{H}_0 = \sum_{a=1}^5 d_a(\mathbf{k}) \,  \Gamma_a   
 + \frac{\mathbf k^2} {2\,m'} \,,
\quad d_a  (\mathbf{k})   = \frac{\tilde d_a(\mathbf{k})   }{2\, m}\,.
 \end{equation}
The $\Gamma_a$ matrices provide a rank-four irreducible representation of the Clifford algebra anticommutation relation $\{\Gamma_a,\Gamma_b\} = 2\, \delta_{ab}$, with an Euclidean (instead of Minkowski) metric. Furthermore,
the functions denoted by $\tilde d_a(\mathbf k)$ are the $\ell =2$ spherical harmonics \cite{Murakami}, and are given by
\begin{align}
\label{ddef}
&\tilde d_1(\mathbf{k}) = \sqrt{3}\, k_y \,k_z\,,
\quad \tilde d_2(\mathbf{k}) =  \sqrt{3}\, k_x\, k_z\, ,\quad
 \tilde d_3(\mathbf{k}) =  \sqrt{3} \,k_x\, k_y\, ,\quad
\tilde d_4(\mathbf{k}) =\frac{\sqrt{3}  \,  (k_x^2 - k_y^2) }{2}\,, \quad
 \tilde d_5(\mathbf{k}) = \frac{2\, k_z^2 - k_x^2 - k_y^2}{2} \,.
\end{align}
Due to the fact that the term ${k^2}/(2m')$ has no spinorial structure, it effectively breaks the particle-hole symmetry in the model, and makes the band masses of the conduction and valence bands unequal. This aspect of the problem will be important for some computations performed in this paper.

The Euclidean action of this system (augmented by $N_f$ fermionic flavors, which allows us to check our calculations using the alternative controlled approximation using the large-$N_f$ limit \cite{Abrikosov,moon-xu}) is given by
\begin{align}
S_0 =&  \int d\tau \,d^3{\mathbf x}
 \left[
 \sum_{i=1}^{N_f}  \psi_i^{\dag}
 \left( \partial_{\tau} + \mathcal{H}_0 + \mathrm{i}\, e 
 \,\varphi \right) \psi_i 
 +\frac{c}{2} \big( \nabla \varphi  \big)^2  \right] ,
\end{align}
where $\psi_i$ denotes the fermionic spinor with flavor index $i$. The Coulomb interaction is mediated by a scalar boson field $\varphi$, that has no dynamics. We note that the physical case is given by $N_f=1$.

If we integrate out $\varphi$ to obtain the Coulomb interaction as an effective four-fermion term, this leads to the following action: 
\begin{align}
S=& \sum_{i=1}^{N_f} 
\int \frac{d\tau \,d^3 {\mathbf k} } {(2\,\pi)^3}\, 
{ \psi}_i^{\dag}(\tau,\mathbf k)
\left(\partial_{\tau} + \mathcal{H}_0 \right)
 { \psi}_i  (\tau, \mathbf k) \nn
& \quad + \frac{e^2}{2 \,c } \sum_{i,j=1}^{N_f}
\int \frac{d\tau\,
d^3 {\mathbf k}\, d^3 {\mathbf k'}\,d^3 {\mathbf q} }
{(2\pi)^{9}}\, 
V(\mathbf q)\,{\psi}^{\dag}_{i} (\tau,{\mathbf k}+\mathbf q)
\,{\psi}_{i}(\tau,{\mathbf k})\,
 {\psi}^{\dag}_{j}( \tau ,{\mathbf k}'-\mathbf q)\, 
\psi_j (\tau,{\mathbf k}')  \,,
\label{action}
\end{align}
where the (instantaneous) Coulomb interaction vertex is simply given by $V(\mathbf q) = 1/\mathbf q^2$ in the momentum space. 

The non-interacting Green's function for each fermionic flavor is given by
\begin{align}
G_0(\omega, \mathbf{k}) =  
\frac{ \mathrm{i}\, \omega- \frac{ \mathbf k^2}{2\,m'}  + \mathbf{d}(\mathbf{k}) \cdot{\mathbf{\Gamma}}}
{-\left ( \mathrm{i}\, \omega- \frac{\mathbf k ^2}{2\, m'}  \right )^2 +|\mathbf{d}(\mathbf{k})|^2}\,,
\label{baregf}
\end{align}
where $|\mathbf{d}(\mathbf{k})|^2 =  \mathbf k^4/(4\,m^2)$. Henceforth, so as not to clutter up the notation, we will use $\mathbf{d}_{\mathbf{k}}$ instead of $\mathbf{d}(\mathbf{k})$.

The controlled NFL fixed point in the LAB phase, obtained via RG analysis of the action in Eq.~\eqref{action}, is described by \cite{moon-xu} 
\begin{align}
\frac{{e^*}^2}{2\,c}= \frac{30\,\pi^2\,\varepsilon} 
{m\,\left(15\,N_f+4\right)}\,,
\end{align}
for $\varepsilon=4-d$ (where $d$ is the number of spatial dimensions).
The dynamical critical exponent $z$ at this fixed point is given by $z^*=2-4\,\varepsilon/(15 \, N_f+4)$.

Using Noether's theorem, the electrical current ($\mathbf{J_c}$) and the momentum ($\mathbf{P}$) operators, associated with the global $U(1)$ symmetry and the continuous spatial translation invariance, respectively, are given by
\begin{align}
\mathbf{J_c}{(t,\mathbf{q})}&=\sum_{i}\int\frac{d^3\mathbf{k}}{(2\,\pi)^3}\,
{\psi}_i^{\dagger}(t,\mathbf{k+q})
\left (\nabla\mathbf{d_k}\cdot\mathbf{\Gamma} \right )
{\psi}_i(t,\mathbf{k}),\quad\\
\mathbf{P}{(t,\mathbf{q})} &=\sum_{i}\int\frac{d^3\mathbf{k}}{(2\pi)^3}
\, (\mathbf{k} + \mathbf{q}/2)\,{\psi}_i^{\dagger}(t,\mathbf{k}+\mathbf{q})\,{\psi}_i(t,\mathbf{k}),
\end{align}
where $t=-\mathrm{i}\,\tau$. Moreover, the Hamiltonian density $h(\mathbf{x})$ can be considered, as a first approximation, to be the heat density of the system. Hence, by using the Fourier-transformed version of the continuity equation
for the heat flow, $\dot{h}(\mathbf{x})+\nabla\cdot \mathbf{J_t}=0$,
the thermal current is given by
\begin{align}
\mathbf{J_{t}}{(t,\mathbf{q})}&=\frac{1}{2}\sum_{i}\int\frac{d^3 \mathbf k}{(2\pi)^3}\left[\partial_t{\psi}_i^{\dagger}(t,\mathbf{k+q})\left(\nabla \mathbf{d}_\mathbf{k}\cdot\mathbf{\Gamma}+\frac{\mathbf{k}}{m'}\right)\psi_i(t,\mathbf{k})\right.
+{\psi}_i^{\dagger}(t,\mathbf{k+q})\left(\nabla \mathbf{d}_\mathbf{k}\cdot\mathbf{\Gamma}+\frac{\mathbf{k}}{m'}\right)\partial_t{\psi}_i(t,\mathbf{k})\bigg],
\end{align}
where $\partial_t{\psi}_i\equiv \mathrm{i}\,[H_0,\psi_i]$, with $H_0$ being the non-interacting Hamiltonian.
In the rest of the paper, we will consider the case with $ N_f = 1$.

\section{Memory matrix formalism}
\label{memory_matrix}

In this section, we will employ the memory matrix approach to calculate the transport properties in the LAB phase of the Luttinger semimetal described by Eq.~\eqref{action}. As mentioned before, this method has the important advantage of not relying on the existence of long-lived quasiparticles in the system \cite{Forster-HFBSCF,Rosch-PRL,Hartnoll-PRB_2013,Freire-AP_2014,Patel-PRB,Hartnoll_PRB_2014,Zaanen-CUP,Lucas-PRB,Freire-AP_2017,Freire-EPL,Sachdev-MIT,Freire-EPL_2018,Berg-PRB,Freire-AP_2020,wang2020low}. 
We will explain here the most important aspects of the application of this formalism, since all the technical details can be found in previous papers by one of the authors  \cite{Freire-AP_2017,Freire-AP_2020}. 

We take into account the fact that the hydrodynamic regime is expected to be the appropriate one to describe the dynamics of the strongly interacting system under study. This is due to the fact that the decay rate associated with the Coulomb interactions is much larger than the electron-impurity scattering rate. In this regime, the only nearly-conserved vector operator in the model (that has a finite overlap with the currents of interest -- to be specified below) turns out to be the total momentum operator itself. Therefore, the matrix of generalized conductivities can be written in a compact form as 
\begin{align}
\label{transp_coeff}
\tilde{\sigma}_{\mathbf{J_i J_i}}(\omega,T)=\frac{\chi^R_{\mathbf{J_iP}}(T)}
{\left  [ M_{\mathbf{PP}}(T)-\mathrm{i} \,\omega \,\chi^R_{\mathbf{P P}}(T)  \right ]
\left [\chi^R_{\mathbf{J_iP}}(T)\right  ]^{-1}}\,,
\end{align}
where the subscript $\mathbf{i}=\{{\mathbf{c,t}}\}$ of the generalized current $\mathbf{J_i}$ refers to the charge current and the thermal current, respectively. The notation $\chi^R_{\mathbf{J_i P}}(T)$ $\left( \chi^R_{\mathbf{P P}}(T) \right) $ corresponds to the static retarded current-momentum (momentum-momentum) susceptibility. Hence, $\chi^R_{\mathbf{J_i P}}(T)$ gives the overlap of the generalized current with the total momentum. The memory matrix $M_{\mathbf{PP}}(T)$ encodes the relaxation mechanism of the nearly-conserved operator (meaning slowly varying) of the theory (i.e., the momentum operator $\mathbf{P}$ in our system), which is relaxed on long timescales. We will further assume, without any loss of generality, that the transport is along the $z$-direction (as all directions are isotropic). Consequently, the current-momentum susceptibility takes the form
\begin{align}
\chi_{\mathrm{J_i^zP_z}}(T)=\int_{0}^{\beta}d\tau 
\left \langle \mathrm{J_i^z(\tau)\, P_z(0)} \right \rangle,
\end{align}
where $\beta=1/T$. To leading order, the memory matrix is given by (again, for transport along the $z$-direction) 
\begin{align}
M_{\mathrm{P_z P_z}}(T) =\int_{0}^{\beta} d\tau 
\left \langle  \mathrm{\dot{P}^{\dagger}_z(0)\,
\frac{\mathrm{i}}
{\omega-L_0}\, \dot{P}_z(\mathrm{i}\,\tau)} \right \rangle \,,
\end{align}
where $\mathrm{L_0}$ is the non-interacting Liouville operator. Finally, the generalized dc conductivity is given by $\sigma_{\mathbf{J_i J_i}}(T)\equiv\tilde{\sigma}_{\mathbf{J_i J_i}}(\omega\rightarrow 0,T)$, which leads to
\begin{align}\label{formal_exp}
\sigma_{\mathrm{J_i J_i}}(T)=\frac{\chi^2_{\mathrm{J_i^z P_z}}(T)}
{  \lim \limits _{ \omega \rightarrow 0}
\frac{\text{Im}\,G^R_{\mathrm{\dot{P}_{z} \dot{P}_{z}}}(\omega,T)}
{\omega} }\,,
\end{align}
where $G^R_{\mathrm{\dot{P}_{z} \,\dot{P}_{z}}}(\omega,T)
=\left \langle \mathrm{\dot{P}_{z}}(\omega) \,
\mathrm{\dot{P}_{z}}(-\omega)\right \rangle_0$ is the retarded correlation function of the momentum operators. The notation $\langle \ldots\rangle_0$ indicates that the average in the grand-canonical ensemble is taken, to leading order, using the non-interacting Hamiltonian of the system.

We now include short-ranged disorder that provides a source of momentum relaxation in the system. We incorporate this by adding an impurity scattering term that couples to the fermionic density in the action as follows:
\begin{align}\label{S_imp}
S_{imp}=\sum_{i}\int d\tau \,d^3\mathbf{x}\, W(\mathbf{x})
\, \psi_i^{\dagger}(\tau, \mathbf{x})\,\psi_i(\tau, \mathbf{x})\,.
\end{align}
Here, we will assume a weak uncorrelated disorder with a Gaussian distribution: $ \langle\langle W(\mathbf{x})\rangle\rangle=0$ and $ \langle\langle W(\mathbf{x})\,W(\mathbf{x'})\rangle\rangle
=W_0^2\, \delta^3(\mathbf{x}-\mathbf{x'})$, where $ W_0^2$ represents the square of the average magnitude of the random potential experienced by the fermions. Therefore, to leading order in $W_0^2$,
we obtain
\begin{align}
\lim_{\omega\rightarrow 0}\frac{\text{Im}\,G^{R}_{\mathrm{\dot{P}_{z} \dot{P}_{z}}}(\omega,T)}{\omega}\approx\lim_{\omega\rightarrow 0}
W_0^2 \int\frac{d^3\mathbf{q}}{(2\pi)^3} \, \frac{\text{Im}\,
\Pi_0^R(\omega, \mathbf{q})}{\omega}
\,,
\end{align}
where $\Pi_0^R(\omega, \mathbf{q})=\Pi_0(\mathrm{i} \,
\omega\rightarrow \omega+ \mathrm{i}\, 0^+, \mathbf{q})$ is the retarded correlation function, with
\begin{align}
& \Pi_0(\mathrm{i} \,\omega, \mathbf{q})
=-T\, \sum_{k_0}\int\frac{d^3\mathbf{k}}{(2\pi)^3}\, k_z^2
\,\text{Tr}
\big[G_0(\mathrm{i} \,k_0+\mathrm{i} \,\omega, \mathbf{k}+\mathbf{q})
\, G_0(\mathrm{i} \,k_0, \mathbf{k})\big]\,.
\end{align}
{We point out that this perturbative treatment of disorder is only valid at intermediate temperature scales. At very low temperatures, this perturbative approach breaks down, since the impurity scattering term of Eq. \eqref{S_imp} turns out to be a relevant perturbation for the LAB phase, and causes a runaway flow to strong disorder under RG \cite{rahul-sid,ips-rahul,*ips-rahul}. With this cautionary remark, we now proceed to calculate the transport coefficients of the LAB phase within such an intermediate temperature regime.}

\begin{figure}[]
\includegraphics[width=0.45\textwidth]{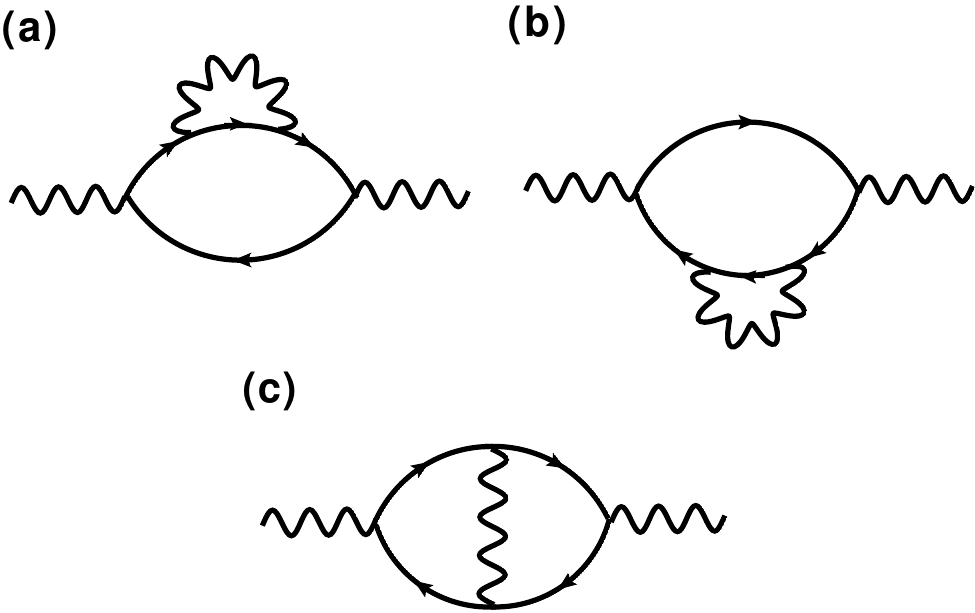}
\caption{\label{diagrams}
Feynman diagrams for the leading-order corrections to the thermal current-momentum susceptibility $\chi_{\mathrm{J_t^{z} P_z}}(T)$, and the electric current-momentum susceptibility $\chi_{\mathrm{J_{c}^z P_z}}(T)$. Sub-figures ${(a)}$ and ${(b)}$ represent the diagrams with self-energy insertions, while sub-figure ${(c)}$ shows the diagram with the vertex correction.} 
\end{figure}

\subsection{Transport coefficients}

In the nearly hydrodynamic regime, the electrical conductivity $\sigma$, the thermal conductivity at zero electric field $\bar{\kappa}$, and the thermoelectric conductivity $\alpha$ are given by the expressions
\begin{align}
\sigma(T)&\equiv\sigma_{\mathbf{J_c J_c}}=\chi^R_{\mathrm{J_c^z P_z}}(T) 
\,M_{\mathrm{P_z P_z}}^{-1}(T)
\,\chi^R_{\mathrm{P_z J_{c}^z}(T)}\,,\label{TC1}\\
\bar{\kappa}(T)&\equiv\frac{\sigma_{\mathbf{J_t J_t}}}{T}
=\frac{1}{T}
\,\chi^R_{\mathrm{J_t^{z} P_z}}(T) \,
M_{\mathrm{P_z P_z}}^{-1}(T)
\, \chi^R_{\mathrm{P_z J_t^{z}}}(T)\,,\label{TC2}\\
\alpha(T)&\equiv\frac{\sigma_{\mathbf{J_c J_t}}}{T}
=\frac{1}{T}\,
\chi^R_{\mathrm{J_c^z P_z}}(T) \,M_{\mathrm{P_z P_z}}^{-1}(T)
\,\chi^R_{\mathrm{P_z J_t^{z}}}(T)\,,\label{TC3}
\end{align}
respectively. The thermal conductivity at zero current is given by
\begin{align}\label{kappa}
\kappa(T)=\bar{\kappa}(T)-\frac{T\,\alpha^2(T)}{\sigma(T)}\,.
\end{align}
Therefore, it is essential to calculate the current-momentum susceptibilities given by $\chi_{\mathrm{J_c^z P_z}}(T)$ and $\chi_{\mathrm{J_t^{z} P_z}}(T)$, and the memory matrix element $M_{\mathrm{P_z P_z}}(T)$, which we will do in the next subsection.

We would like to point out that the thermoelectric properties are usually measured using the Seebeck coefficient $S$, which is given by the ratio of the thermoelectric conductivity and the electrical conductivity (i.e., $S=\frac{\alpha} {\sigma}$). If the model is symmetric under particle-hole transformation, the thermoelectric coefficient $\alpha$ will vanish identically. As a result, in order to achieve a finite thermoelectric response, breaking the particle-hole symmetry is crucial on physical grounds.

To further illustrate the point mentioned above, we note that for equal valence and conduction band masses (i.e., in the limit of $m' \rightarrow \infty $), the electric current-momentum susceptibility $\chi_{\mathrm{J_c^z P_z}}$ is completely suppressed at zeroth order, in view of the fact that only odd powers of $k_0$ appear in the numerator after performing the trace of the integrand. At next-to-leading order in the perturbation theory, the contributions to $\chi_{\mathrm{J_c^z P_z}}$ due to self-energy insertions (represented by the sum of diagrams in Figs.~\ref{diagrams}(a) and \ref{diagrams}(b)) are given by
\begin{align}
& \chi_{\mathrm{J_c^z P_z}}^{(\mathrm{a+b})} (T)
= 
  -\left(\frac{e^2}{c} \right)
 T^2 \sum \limits_{k_0,\ell_0}
\int \frac{d^3 {\mathbf k} \, d^3 {\boldsymbol \ell}}{(2\pi)^6} \, k_z \,
\text{Tr}
\left [ \left(\partial_{k_z}
\mathbf{d}(\mathbf{k}) \cdot \mathbf{\Gamma} \right) 
G_0(k_0, \mathbf{k})
\frac{G_0( \ell_0,  \mathbf{k} + \boldsymbol{\ell})} {\boldsymbol {\ell}^2}
G_0(k_0, \mathbf{k} )
\,G_0(k_0, \mathbf{k}) \right] \nn
& =   -\left(\frac{e^2}{c} \right)
 T^2\sum \limits_{k_0,\ell_0}
\int \frac{d^3 {\mathbf k}\, d^3 {\boldsymbol \ell}}{(2\pi)^6}\, k_z \,
\frac{ \text{Tr}
\left [ \left(\partial_{k_z}
\mathbf{d}(\mathbf{k}) \cdot \mathbf{\Gamma} \right)   
\frac{ \mathrm{i}\, k_0+ \mathbf{d}(\mathbf{k}) \cdot{\mathbf{\Gamma}}}   
{\left ( \mathrm{i}\,k_0  \right )^2
-| \mathbf{d}(\mathbf{k})|^2} 
\, \frac{  \mathrm{i}\,\ell_0+ \mathbf{d}( \mathbf{k} ) \cdot{\mathbf{\Gamma}}} 
 {
\left ( \mathrm{i}\,\ell_0  \right )^2
- | \mathbf{d}( \mathbf{k} )|^2}
\, 
\frac{ \mathrm{i}\, k_0+ \mathbf{d}(\mathbf{k}) \cdot{\mathbf{\Gamma}}}   
{\left ( \mathrm{i}\,k_0  \right )^2
-| \mathbf{d}(\mathbf{k})|^2}
\,\frac{ \mathrm{i}\, k_0+ \mathbf{d}(\mathbf{k}) \cdot{\mathbf{\Gamma}}}   
{\left ( \mathrm{i}\,k_0  \right )^2
-| \mathbf{d}(\mathbf{k} )|^2} \right] }
{ \left (\mathbf k-\boldsymbol {\ell} \right ) ^2}\nn
&= 0
\,.
\end{align}
These contributions vanish due to the same reason as before: they contain either odd powers of $k_0$ or odd powers of $\ell_0$ in the numerator after performing the trace in the integrand. 
One can straightforwardly check that a similar result holds for the vertex correction shown in Fig.~\ref{diagrams}(c). In fact, this null result for $\chi_{\mathbf{J_c P}}$ holds at all higher-order loops. This is due to particle-hole symmetry, which is present for equal valence and conduction band masses at zero doping. Since the operators $\mathbf {J_c}$ and  $\mathbf P$ are odd and even, respectively, under particle-hole symmetry, their overlap (i.e., the electric current-momentum susceptibility $\chi_{\mathbf{J_c P}}$) must vanish at all loop orders. This implies that both $\alpha$ and $\sigma$ are suppressed under this particular condition.

We point out that the vanishing of $\alpha$ and $\sigma$ is no longer true for a finite $m'$. Therefore, we will henceforth assume a broken particle-hole symmetry, corresponding to a finite $m'$ in Eq.~\eqref{bare}, in all calculations that follow. We emphasize that it is crucial for the Luttinger semimetal compounds to possess such a broken symmetry for the transport phenomena calculated in this work to be observed experimentally. We also point out that a large particle-hole asymmetry is indeed observed in semimetallic HgTe compounds, which leads to a finite thermoelectric response in these materials \cite{Markov_2018}.

\subsection{Thermal and electric current-momentum susceptibilities at finite $T$}

\begin{figure}[]
\includegraphics[width=0.44\textwidth]{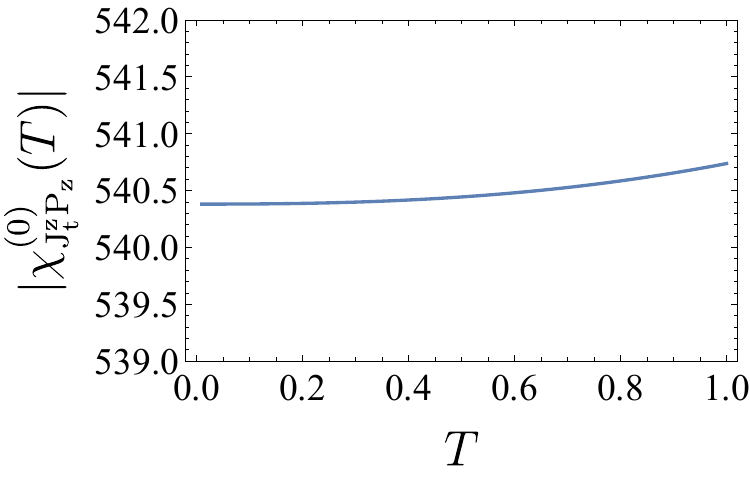}
\caption{\label{Fig:chi_QP}
Numerical plot of the thermal current-momentum susceptibility at zeroth order ($\chi^{(0)}_{\mathrm{J_t^z P_z}}(T)$) versus temperature $T$. Here, we have chosen the parameters $m=1$, $m'=5$, $N_f=1$, and $\Lambda=10$. The temperature dependence of this one-loop contribution is found to be $|\chi^{(0)}_{\mathrm{J_t^z P_z}}(T)|\sim 540.4+0.4\,T^{5/2}$. Since the prefactor associated with the $T$-independent term is always much larger (by some orders of magnitude) than the one associated with the $T$-dependent term, $\chi^{(0)}_{\mathrm{J_t^z P_z}}(T)$ will effectively have no appreciable temperature dependence.} 
\end{figure}

We first calculate the thermal current-momentum susceptibility at zeroth order, which is given by
\begin{align}
\label{chiJP}
\chi^{(0)}_{\mathrm{J_t^z P_z}}(T)
& =-\lim_{\mathbf{q} \rightarrow 0}  T \,
\sum \limits_{k_0}
\int \frac{d^3 {\mathbf k}}{(2\pi)^3} \,\mathrm{i}\,k_0\,k_z 
\,\text{Tr}
\left [ \left(\partial_{k_z}
\mathbf{d}_{\mathbf{k}} \cdot \mathbf{\Gamma}+\frac{{k_z}}{m'} \right)  G_0(k_0, \mathbf{k}+\mathbf{q})
\, G_0(k_0, \mathbf{k}) \right].
\end{align}
To perform the summation over the fermionic Matsubara frequency $k_0$, we apply the method of residues. We then proceed to solve Eq.~\eqref{chiJP} by means of numerical techniques, which gives $\chi^{(0)}_{\mathrm{J_t^z P_z}}(T)\sim \mathcal{A}+\mathcal{B}\,T^{5/2}$, as depicted in Fig.~\ref{Fig:chi_QP}. Following a careful analysis, we find that the prefactor $\mathcal{A}$ is always much larger than $\mathcal{B}$ by several orders of magnitude, implying that $\chi^{(0)}_{\mathrm{J_t^z P_z}}(T)$ has no appreciable temperature dependence. Therefore, one can approximate it by a $T$-independent constant, i.e., $\chi^{(0)}_{\mathrm{J_t^z P_z}}(T)\sim \mathcal{A} $.

\begin{figure}[th]
\includegraphics[width=0.41\textwidth]{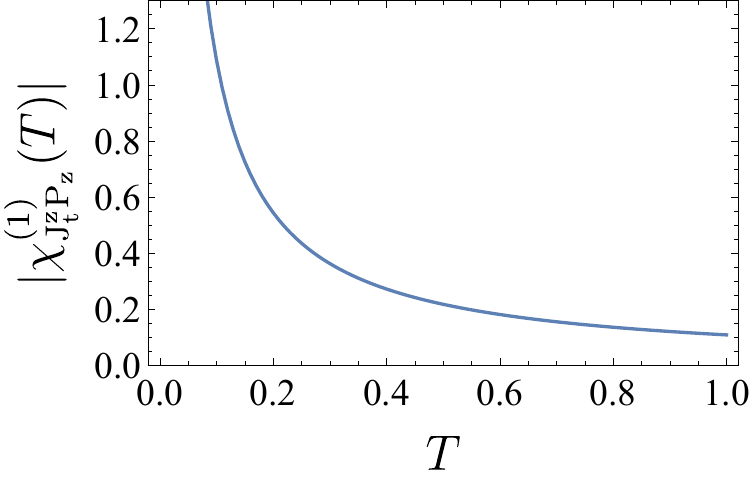}
\caption{\label{Fig:chi_QP_2loops}
Numerical plot of the leading-order correction ($\chi^{(1)}_{\mathrm{J_t^z P_z}}(T)$) to the thermal current-momentum susceptibility versus temperature $T$. Here, we have chosen the parameters $m=1$, $m'=5$, $N_f=1$, $e=0.1$, $c=1$, and $\Lambda=10$. The temperature dependence of this contribution is found to be given by $|\chi^{(1)}_{\mathrm{J_t^z P_z}}(T)|\sim 0.1/T$.} 
\end{figure}

For computing the leading order correction due to the Coulomb interactions within the perturbation theory, we note that there are three Feynman diagrams, as shown in Fig~\ref{diagrams}. These contributions evaluate to $\chi^{(1)}_{\mathrm{{J_t^z P_z}}}(T)=\chi^{\mathrm{(a+b)}}_{\mathrm{J_t^z P_z}}(T)+\chi^{\mathrm{(c)}}_{\mathrm{J_t^z P_z}}(T)$, where
\begin{align}
\chi^{\mathrm{(a+b)}}_{\mathrm{J_t^z P_z}}(T)\sim 
\left(\frac{8\,e^2 \Lambda}
{15\, \pi^3 c\, T}\right)
T\,\sum \limits_{k_0}
\int \frac{d^3 {\mathbf k}}{(2\pi)^3}\, &\left\{\mathrm{i}\,k_0\,k_z\frac{\left[\left (\mathrm{i}\,k_0-\frac{\mathbf{k}^2}{2m'}  \right )^3
\left( \partial_{k_z}\mathbf{d_k}\cdot \mathbf{d_k} \right)+3\left (\mathrm{i}\,k_0-\frac{\mathbf{k}^2}{2m'}  \right )
\left( \partial_{k_z}\mathbf{d_k}\cdot \mathbf{d_k}\right) |\mathbf{d_k}|^2 \right]}{\left[\left (\mathrm{i}\,k_0-\frac{\mathbf{k}^2}{2m'}  \right )^2-|\mathbf{d_k}|^2\right]^3}\right.\nonumber\\
&\left.
\quad + \left(\frac{\mathrm{i}\,k_0\,k_z^2}{m'}\right)\frac{3\left (\mathrm{i}\,k_0-\frac{\mathbf{k}^2}{2m'}  \right )^2  |\mathbf{d_k}|^2 +|\mathbf{d_k}|^4
}{\left[\left (\mathrm{i}\,k_0-\frac{\mathbf{k}^2}{2m'}  \right )^2-|\mathbf{d_k}|^2\right]^3}\right\}\,,
\label{ex1}
\end{align}
\begin{align}
&\chi^{\mathrm{(c)}}_{\mathrm{J_t^z P_z}}(T)
\sim \left(\frac{e^2\,\Lambda}{2\, \pi^3 c\, T}\right)
T\,\sum \limits_{k_0}
\int \frac{d^3 {\mathbf k}}{(2\pi)^3}\,\left\{\frac{\mathrm{i}\,k_0\, k_z\left (\mathrm{i}\,k_0-\frac{\mathbf{k}^2}{2m'}  \right )
\left( \partial_{k_z}\mathbf{d_k}\cdot \mathbf{d_k} \right)}
{\left[\left (\mathrm{i}\,k_0-\frac{\mathbf{k}^2}{2m'}  \right )^2-|\mathbf{d_k}|^2\right]^2}+
\left(\frac{\mathrm{i}\,k_0\,k_z^2}{2m'}\right)\frac{\left (\mathrm{i}\,k_0-\frac{\mathbf{k}^2}{2m'}  \right )^2+|\mathbf{d_k}|^2}{\left[\left (\mathrm{i}\,k_0-\frac{\mathbf{k}^2}{2m'}  \right )^2-|\mathbf{d_k}|^2\right]^2}\right\}\,,
\label{ex2}
\end{align}
with $\Lambda$ being the ultraviolet cutoff scale. The details for the calculation of the above expressions are provided in Appendix \ref{leading_corrections}. Next, we solve the summations over the Matsubara frequency $k_0$ analytically, and evaluate the resulting integrals using the standard numerical integration methods. Finally, we find that
$\chi^{(1)}_{\mathrm{{J_t^z P_z}}}(T)\sim (e^2/c)\,T^{-1}$, as shown in Fig.~\ref{Fig:chi_QP_2loops}.

The electric current-momentum susceptibility as a function of temperature has been calculated by us in an earlier paper \cite{freire_2020}. Nevertheless, for the sake of completeness, we will briefly review the results here. At zeroth order, this susceptibility is given by
\begin{align}
\label{chiJP2}
\chi^{(0)}_{{\mathrm{J_c^z P_z}}}(T)
& =-\lim_{\mathbf{q} \rightarrow 0} T \,
\sum \limits_{k_0}
\int \frac{d^3 {\mathbf k}}{(2\pi)^3} \,k_z 
\,\text{Tr}
\left [ \left(\partial_{k_z}
\mathbf{d}(\mathbf{k}) \cdot \mathbf{\Gamma} \right)  G_0(k_0, \mathbf{k}+\mathbf{q})
\, G_0(k_0, \mathbf{k}) \right] \,.
\end{align}
By solving the right-hand-side, we have obtained the temperature dependence to be $\chi^{(0)}_{{\mathrm{J_c^z P_z}}}(T)\sim T^{3/2}$.
The leading-order corrections due to Coulomb interactions are given by diagrams similar to Fig.~\ref{diagrams}.  Solving the corresponding loop integrals, we have found that $\chi^{(1)}_{{\mathrm{J_c^z P_z}}}(T)\sim (e^2/c)\,T^{1/2}$.

\subsection{Memory matrix calculation}

\begin{figure}[t]
\includegraphics[width=0.15\textwidth]{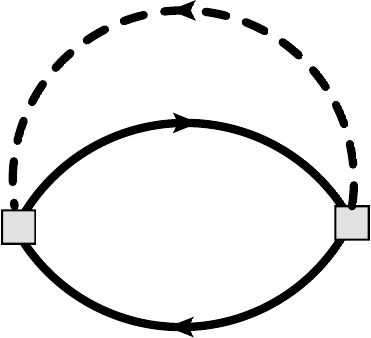}
\caption{\label{Fig:m_0}
Feynman diagram for the computation of the leading-order contribution $M^{(0)}_{\mathrm{P_z P_z}}(T)$ to the memory matrix. The solid line stands for the bare fermionic propagator of the model, whereas the dashed line denotes the impurity line that results in an effective four-fermion interaction.} 
\end{figure}

\begin{figure}[t]
\includegraphics[width=0.43\textwidth]{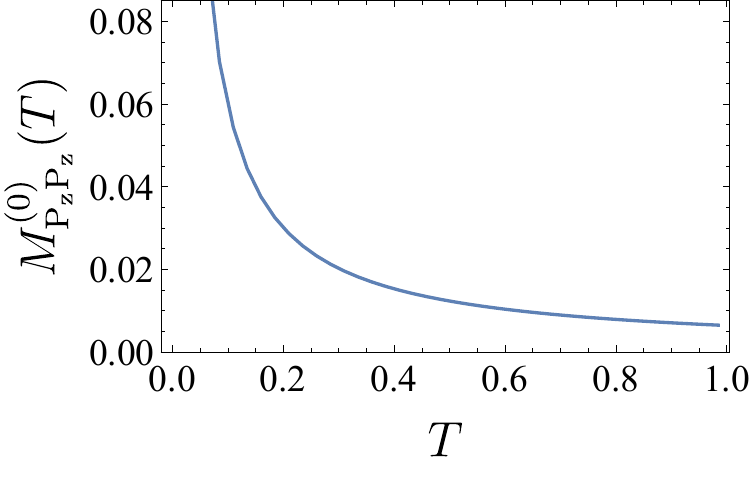}
\caption{\label{Fig:M_PP}
Numerical plot of the memory matrix $M^{(0)}_{\mathrm{P_z P_z}}(T)$ versus temperature $T$. Here, we have chosen the parameters $W_0=1$, $N_f=1$, $m=1$, $m'=5$, and $\Lambda=10$. To compute $M^{(0)}_{\mathrm{P_z P_z}}$, we have performed an analytical continuation $\mathrm{i}\,\omega\rightarrow \omega+\mathrm{i}\,\delta$ by setting $\delta = 10^{-8}$. The curve corresponds to the fit given by $f(T)=\mathcal{C}+\mathcal{D}/T$, where the parameters are $\mathcal{C}\approx 4.7\times10^{-4}$ and $\mathcal{D}\approx 5.9\times 10^{-3}$.} 
\end{figure}

We now calculate the memory matrix for the model. As explained before, in the strongly interacting LAB phase, only the momentum operator is associated with a small relaxation rate. Therefore, $\mathbf P$ is the only nearly-conserved operator in the transport theory. Fig.~\ref{Fig:m_0} shows the Feynman diagram corresponding to the memory matrix calculation to leading order in $W_0^2$, which evaluates to
\begin{align}
M^{(0)}_{\mathrm{P_z P_z}}(T)&=-W_0^2 \,
\lim_{\omega\rightarrow 0}\frac{\text{Im}
\bigg [
\int \frac{d^3 {\mathbf k} \, d^3 {\mathbf q}}{(2\pi)^6}
\,k_z^2\, T\sum \limits _{k_0}\text{Tr}\left[G_0(\omega+k_0,\mathbf{k+q})\,
G_0(k_0,\mathbf{k})\right]  \bigg ]
\Bigg|_{\mathrm{i} \,\omega\rightarrow\omega+\mathrm{i} \,\delta}}
{\omega}
\nonumber\\
&=-4\,W_0^2\,
\lim_{\omega\rightarrow 0}\frac{\text{Im}
\bigg[
\int \frac{d^3 {\mathbf k} \, d^3 {\mathbf q}}{(2\pi)^6}\, k_z^2\, 
T\sum \limits _{k_0}
\frac{\left \lbrace \mathrm{i} \,k_0+\mathrm{i} \,\omega
-\frac{ \left(\mathbf{k+q} \right )^2}{2m'}\right \rbrace
\left(\mathrm{i} \,k_0-\frac{\mathbf k^2}{2m'}\right)
+ \left( \mathbf{d}_{\mathbf{k+q}}\cdot\mathbf{d}_{\mathbf{k}} \right)
}
{\left \lbrace \left(\mathrm{i} \,k_0+\mathrm{i} \,\omega
-\frac{(\mathbf{k+q})^2}{2m'}\right)^2-|\mathbf{d}_{\mathbf{k+q}}|^2\right \rbrace
\left \lbrace \left(\mathrm{i} \,k_0
-\frac{\mathbf k^2}{2m'}\right)^2-|\mathbf{d}_{\mathbf{k}}|^2
\right \rbrace }\bigg ]
\Bigg|_{\mathrm{i} \,\omega\rightarrow\omega+\mathrm{i} \,\delta}}{\omega}
\,.
\end{align}
The summation over $k_0$ is performed analytically, using the method of residues.
The resulting integral is then evaluated numerically, which yields $M^{(0)}_{\mathrm{P_z P_z}}(T)/W_0^2\sim \mathcal{C}+\mathcal{D}/T$ (see Fig. \ref{Fig:M_PP}), where $\mathcal{C}$ and $\mathcal{D}$ are non-universal constants that depend on the ultraviolet cutoff $\Lambda$ of the model. In fact, it can be shown numerically that $\mathcal{C}$ scales as $\Lambda^{2}$, while $\mathcal{D}$ scales as $\Lambda^4$, which lead to $\frac{\mathcal{C}} {\mathcal{D}}\rightarrow 0$ for $\Lambda\rightarrow \infty$. Therefore, the memory matrix element at low temperatures can be approximated as $M^{(0)}_{\mathrm{P_z P_z}}(T)/W_0^2\approx \mathcal{D}/T$. We will use this result in what follows.

\section{Results}
\label{results}

Using Eqs.~\eqref{TC1}-\eqref{TC3}, we are now ready to calculate the transport coefficients of the LAB phase within the memory matrix formalism. At zero doping and low temperatures, these coefficients obey
\begin{align}\label{thermoelectric2}
\bar{\kappa}(T)&\sim T^{-n}, \text{ with } 0\lesssim n \lesssim 1\,, \\
\alpha(T)&\sim T^{p}, \text{ with } 1/2\lesssim p \lesssim 3/2\,.
\label{thermoelectric}
\end{align}
We point out that the main effect of the Coulomb interactions is to increase the value of the exponent $n$, and to decrease the magnitude of the exponent $p$, thereby enhancing $\bar{\kappa}$ and $\alpha$ at low temperatures ($T<1$).
The electrical conductivity $\sigma(T)$, calculated by us earlier \cite{freire_2020}, indicates that at zero doping and low temperatures,
$\sigma(T)\sim T^{n'}$ (for $2\lesssim n'\lesssim 4$).

Given all the results above, we conclude that, even though the Luttinger semimetal is a power-law insulator at zero doping, the thermal conductivity at zero electric field can in fact be very high at low-enough temperatures. Despite this statement, experimentally speaking, the physical quantity that is usually measured in the context of heat transport turns out to be the thermal conductivity at zero current $\kappa$, which is instead given by Eq.~(\ref{kappa}). 

At this point, it is also interesting to analyze the status of the Wiedemann-Franz (WF) law in the LAB phase. The WF law in a Fermi liquid regime (with well-defined quasiparticles), in the presence of weak disorder, is universally given by the Lorenz ratio $\mathcal{L}=\kappa/(\sigma \, T)=\pi^2/3$ at low temperatures (the Boltzmann constant $k_B$ and the elementary charge $e$ are set equal to unity, for simplicity). Large violations of the WF law (see, for example, Ref.~\cite{Lucas_PNAS} for discussions in the context of a Weyl semimetal) are expected in a system with no long-lived quasiparticles, such as in the LAB phase. Indeed, from Eq.~\eqref{kappa}, we obtain that $\kappa$ vanishes to leading order, i.e., $\mathcal{L}$ turns out to be dramatically suppressed in this case {(see Appendix \ref{estimate} for a more detailed discussion)}. Therefore, to leading order, no significant electronic contribution to $\kappa$ is expected. We emphasize that the contributions from the phonons, which are of course present in any material, have not been considered in our analysis here. The phonon-contribution is naturally expected to dominate the thermal transport of the system at high-enough temperatures (above the Debye temperature). 

Finally, as mentioned earlier, another interesting transport coefficient that provides useful information about the system is the thermoelectric conductivity, measured by the Seebeck coefficient $S$. From Eq. \eqref{thermoelectric}, we conclude that $S$ can also be quite large in the LAB phase. This implies that the thermoelectric efficiency, captured by the figure-of-merit $z\,T=S^2\,\sigma\, T/\kappa$, can also be very high in these materials {(see Appendix \ref{estimate})}. This indicates that Luttinger semimetals might have the potential of being ideally suited for many thermoelectric applications. Recent experimental data indeed confirms this -- for example, semimetallic HgTe compounds are potentially good candidates for achieving high figure-of-merit \cite{Markov_2018}. According to our calculations, other closely-related materials with stronger correlations (such as the pyrochlore iridates \cite{Witczak-Krempa}, among other compounds) are expected to be even more efficient for such applications.

\section{Summary and outlook}
\label{conclusions}

In this paper, we have performed a transport calculation of the thermoelectric and thermal properties for the LAB phase in the nearly-hydrodynamic regime, where the Coulomb interactions are much stronger than the coupling of the electrons to quenched disorder. We have applied the memory matrix method for these computations, because this formalism does not depend on the existence of well-defined quasiparticles at low energies. We have calculated the thermal conductivity $\bar{\kappa}$ at zero electric field, and have shown that $\bar{\kappa}\sim T^{-n}$ (with $0\lesssim n\lesssim 1)$ at low temperatures. In addition, we have also computed the thermoelectric coefficient $\alpha$, and have found that $\alpha\sim T^{p}$ (with $1/2\lesssim p\lesssim 3/2)$. These starkly unusual behaviors of the transport coefficients turn out to be the defining features for the existence of this long sought-after NFL phase in Luttinger semimetals. Moreover, we have also proposed that Luttinger semimetals might be another viable candidate for achieving high figure-of-merit for thermoelectric applications.

{Finally, we would like to point out that it will be interesting to see if other strategies for the computation of transport properties, such as the quantum Boltzmann equation approach or the Kubo formula, are able to reproduce the thermal and thermoelectric conductivities due to weak-disorder effects, which are obtained here using the memory matrix approach. In addition, we} would also like to emphasize that, even though there are still no established experimental results to date regarding these transport coefficients (to the best of our knowledge), we hope that our theoretical predictions will stimulate more experimental work on these very interesting strongly correlated materials.

\section*{Acknowledgments}

H.F. acknowledges funding from CNPq under Grant No. 310710/2018-9. 


\appendix

\section{Leading-order corrections to the thermal current-momentum susceptibility}
\label{leading_corrections}

For the leading-order corrections represented by the diagrams with self-energy insertions [see Fig.~\ref{diagrams}(a) and Fig.~\ref{diagrams}(b)], we have
\begin{align}
\label{chiJP1}
& \chi^{\mathrm{(a+b)}}_{\mathrm{J_t^z P_z}}(T)
= - 2\,T\,
\sum \limits_{k_0}
\int \frac{d^3 {\mathbf k}}{(2\pi)^3}\,  
\mathrm{i}\,k_0\,k_z \,
 \text{Tr}\bigg[ 
 \bigg (  \partial_{k_z}\mathbf{d}_{\mathbf k}\cdot \mathbf \Gamma 
+ \frac{k_z}{m'}\bigg) \, G_0(k_0,\mathbf{k})
\, \Sigma(k_0,\mathbf{k}) \,G_0^2(k_0,\mathbf{k})   \bigg]\,,
\end{align}
with
\begin{align}
\label{chiJP1-2}
\Sigma(k_0,\mathbf{k})
& =-\frac{e^2}{2\,c}\,T
\sum \limits _{\ell_0}\int \frac{d^3 {\boldsymbol \ell}}{(2\pi)^3}
\frac{G_0(k_0+\ell_0,\mathbf k+\boldsymbol \ell)}
{\boldsymbol \ell^2}
=
-\frac{e^2}{15 \, \pi^3 c}\left[\frac{\Lambda}{T}
\left ( \mathbf{d}_{\mathbf k}\cdot \mathbf \Gamma\right )
-\frac{5\,m}{4\,\Lambda_{\mathrm{IR}}}\,\big|\mathbf{d}_{\mathbf k} \big | \right],
\end{align}
where $\Lambda$ and $\Lambda_{\mathrm{IR}}$ refer to the ultraviolet and infrared cutoff scales, respectively.
Performing the trace in Eq.~\eqref{chiJP1}, we get the expression in Eq.~\eqref{ex1}.
To obtain the leading-order $T$-dependent scaling behavior of $\chi^{\mathrm{(a+b)}}_{\mathrm{J_t^z P_z}}(T)$, we neglect the $T$-independent term in Eq.~\eqref{chiJP1-2}.

For the two-loop diagram with vertex corrections depicted in Fig.~\ref{diagrams}(c), we have
\begin{align}
&  \chi^{\mathrm{(c)}}_{\mathrm{J_t^z P_z}}(T) = - T\,
\sum \limits_{k_0}
\int \frac{d^3 {\mathbf k}}{(2\pi)^3}\,  
\mathrm{i}\,k_0\,k_z \,\text{Tr}
\bigg[ \bigg(  \partial_{k_z}\mathbf{d}_{\mathbf k}\cdot \mathbf \Gamma
+ \frac{k_z}{m'}\bigg) \,G_0(k_0,\mathbf{k})\,
\Pi_1(k_0,\mathbf{k})\, G_0(k_0,\mathbf{k}) \bigg]\,,
\end{align}
where 
\begin{align}
\Pi_1(k_0,\mathbf{k}) & =-\frac{e^2}{c}\,T\,
\sum \limits_{\ell_0}\int \frac{d^3 {\boldsymbol \ell}}{(2\pi)^3}
\frac{G_0^2(k_0+\ell_0,\mathbf k+\boldsymbol \ell)\,
}
{\boldsymbol \ell^2}
=-\frac{e^2}{16 \, \pi^3 c\, T}\left(\Lambda+\frac{2\,m}{3\,\Lambda_{\mathrm{IR}}}
\,\big |\mathbf{d}_{\mathbf k} \big |\right).
\label{chiJP2-2}
\end{align}
Performing the trace in the integrand, we get the expression in Eq.~\eqref{ex2}.
To obtain the leading-order $T$-dependence, we neglect the second term in Eq~\eqref{chiJP2-2}.

\section{Approximate analysis of the $T$-dependence of the Lorenz ratio and the figure-of-merit}
\label{estimate} 

In this appendix, we provide an approximate analysis of the Lorenz ratio $\mathcal{L}$ and the figure-of-merit $z\,T$ discussed in the main text at low-enough temperatures. For doing so, we consider a very simple approximation in which only the leading-order contributions in $T$, that we calculated in Eqs. \eqref{thermoelectric2} and \eqref{thermoelectric}, are taken into account. We point out that this approximation is only a rough estimate and has to be taken with a grain of salt, because, as we explained before, our perturbative treatment of disorder is not allowed to go to very low temperatures in the LAB phase. 

Within our approximation, the $T$-dependence of the transport coefficients are given by the following power-law scalings: $\sigma\sim T^2$, $\bar{\kappa}\sim T^{-1}$, $\alpha\sim T^{1/2}$ and $\kappa\sim u\,T^{-1}$, where $u$ is a prefactor that is parametrically small (i.e., $u\ll 1$), which reflects the suppression of $\kappa$ obtained from Eq. \eqref{kappa} within the nearly-hydrodynamical regime. Hence, we conclude that the temperature dependence of the Lorenz ratio and the figure-of-merit are given by $\mathcal{L}(T)\propto u/T^4$ and $z\, T\propto T/u$, respectively. It is interesting to note that even though $\mathcal{L}$ seems to show a tendency to be strongly enhanced at low-enough temperatures, the prefactor $u$ will keep its numerical value necessarily equal to a very small quantity, in agreement with our discussion presented in the main text. Likewise, although the scaling of $z\, T$ suggests that this quantity might have a qualitative tendency to be linearly suppressed at low temperatures, the small prefactor $u$ will push it to a very large value, which is also consistent with the analysis presented in the main text.

\bibliography{biblio}

\end{document}